\documentclass[12pt,pictex,psfig,epsf]{article}
\usepackage{epsfig}
\newcommand{\bqa}{\begin{eqnarray}}
\newcommand{\eqa}{\end{eqnarray}}
\newcommand{\beq}{\begin{equation}}
\newcommand{\eeq}{\end{equation}}
\def\square{\vcenter{\vbox{\hrule height.4pt
          \hbox{\vrule width.4pt height8pt
          \kern8pt\vrule width.4pt}\hrule height.4pt}}}

\date{}

\title{\Large \bf The Screening Mass Squared in Hot Scalar Theory\\
to Order $g^5$ Using Effective Field Theory }
\author{\normalsize Jens O. Andersen, \\
\normalsize Department of Physics,\\
\normalsize University of Oslo, \\
\normalsize P.O. BOX 1048, Blindern,\\
\normalsize N-0316 Oslo, Norway
\thanks{Address after September 1, 1997: Ohio State University, Department of
Physics, 174 West 18th Avenue, Columbus, OH-43210-1106, USA.}}

\begin{document}
\maketitle
\begin{abstract}
{\footnotesize We study massless $g^2\Phi^4$-theory at high temperature
and with zero chemical potential. Using modern effective field
theory methods,
we derive the screening mass squared to order $g^5$. 
It is demonstrated that the result is renormalization group invariant
to this order in the coupling constant. 
Renormalization group methods are used to sum up leading logarithms of $g$.
}\\ \\ 
PACS number(s): 11.10.Wx, 12.20.DS, 12.38.Bx 
\end{abstract}
\small
\normalsize
\section{Introduction}
There has been tremendous progress in perturbative calculations in thermal
field theory, since the work on symmetry behaviour at finite temperature
by Dolan and Jackiw more than two decades ago~\cite{dolan}.

One major breakthrough was the discovery by Braaten and Pisarski that 
ordinary perturbation theory breaks down high temperature for soft external
momenta $k$ ($k\sim gT$, where $g$ is the coupling constant)~\cite{pis}. 
The usual connection between the number of loops and powers of the coupling
constant is lost. The leading order results for
physical quantities receive contribution from all
orders in perturbation theory. The solution to the problem is
to use an effective expansion which resums this infinite subset of diagrams
and where loop corrections are truly perturbative (down by powers of $g$).
This resummation program and extensions thereof, have been applied to 
calculate a variety of static and dynamical quantities in hot plasmas
(see e.g. Ref. [3,4]).
The literature is now vast, but we shall confine ourselves to 
briefly mention some papers relevant for the present work.
The electric screening masses squared of SQED and QED have been
calculated by Blaizot {\it et al.}~\cite{parw1} 
to order $e^4$ and $e^5$, respectively. 
The plasmon mass in $g^2\Phi^4$-theory
has been calculated by Parwani to order $g^4$ in 
resummed perturbation theory~\cite{plas}. 
The pressure 
has also 
been computed to order $g^4$ by Frenkel, Saa and Taylor~\cite{frenkel},
and extended to order $g^5$ by Parwani and Singh~\cite{parw3}.
Parwani and Corian\`o have carried out corresponding calculations
in QED to order $e^4$~\cite{par2}, 
and the calculational frontier has again be pushed 
to order $e^5$ by Parwani~\cite{parw2}. In QCD, Arnold and Zhai have obtained
the free energy to order $g^4$~\cite{arnold1}, while Zhai and Kastening
have performed a $g^5$-calculation~\cite{kast}.

Another major achievement is the application of effective field theory
methods to the calculations of static quantities. 
The general idea is to take advantage of two or more well separated 
mass scales in the problem at hand, and construct a sequence of
effective field theories
which are valid at successively longer distances.
The short-distance effects are then
encoded in the parameters of the effective 
Lagrangian~[13-15]. 
In the case of $g^2\Phi^4$-theory
we have the scale $T$ which is associated with the nonzero Matsubara 
frequencies (their masses are $2\pi nT$, where $n$ is a positive integer)
and the scale $gT$ which is associated with the zero-frequency mode (this
mode acquires a thermal mass of order $gT$. Alternatively, this is the
scale at which static scalar fields are screened).
The strategy is to integrating out the scale $T$, or equivalently
to integrating out the heavy modes, to obtain an effective Lagrangian
of the light mode. The effective theory is three dimensional and the
process is called dimensional reduction.
In Non-Abelian gauge theories there are three momentum scales, 
namely $T$ which again is associated with the nonstatic modes, 
$gT$ which is the scale of colour electric screening
and the scale $g^2T$ which is the scale of colour magnetic screening.
In this case it proves useful to construct a second effective field
theory by integrating out the scale $gT$~\cite{br2}.

The methods of effective field theory
were first applied to high temperature field theory 
by Ginsparg and Landsman in Refs.~[17,18].
In these papers, effective three-dimensional Lagrangians were constructed
by explicitly integrating out heavy modes at the one-loop level.
Later, Kajantie {\it et al.} have extended this method beyond the one-loop
level by matching Greens functions in the full and in the effective theory, 
using the effective potential~\cite{laine}. 
This approach has mainly be used in connection
with the study of phase transitions in the standard model and other 
theories [19,20].

There is a nice alternative to explicitly distinguishing between light
and heavy modes. Instead, one writes down the most general effective 
three-dimensional Lagrangian
which is consistent with the symmetries at high temperature. 
The parameters in the effective theory are then determined by the
requirement that static correlators in the full theory are reproduced to
some desired accuracy by the effective theory, at long distances $R\gg 1/T$.
This approach has been developed by Braaten and Nieto in Ref.~\cite{braaten}, 
where they
applied it to $g^2\Phi^4$-theory. Combining this with renormalization group
methods, they computed the screening mass squared
to order $g^5\ln g$ and  the pressure to order $g^6\ln g$.
Later, Braaten and Nieto computed the free energy in QCD, through
order $g^5$~\cite{braaten2}, 
and confirmed the resummation results of Zhai and 
Kastening~\cite{kast}. 
The method has also been applied by the present author to calculate
the screening mass in SQED and QED to order $e^4$ and $e^5$, respectively, 
as well as the free energy in QED to order $e^5$~\cite{jens}. 
These calculations
have reproduced the results first obtained by resummation~[5,9,10]. 

In this letter we apply these ideas to scalar theory and calculate the
screening mass squared to order $g^5$. Using the renormalization group
techniques of Ref.~\cite{braaten}, we also sum up leading logarithms 
of $T/(gT)$ from
higher order of perturbation theory. In section two, we briefly discuss
the effective three-dimensional theory and we determine the
coefficients in the effective Lagrangian. In section three, we calculate
the screening mass to order $g^5$. In section four we summarize our results
and in the appendices A and B, the notation and conventions are given. 
We also
list the sum-integrals of the full theory
and the integrals of the effective theory, which are necessary for the present 
calculations.
\section{The Effective Lagrangian}
In this section we discuss the construction of 
the effective three-dimensional Lagrangian and we determine the
parameters to the accuracy necessary for calculating the
screening mass squared to order $g^5$. Since the results in this
section appear elsewhere
in the literature [18,21], the discussion will 
be rather brief. In particular, we
refer to the 
paper by Braaten and Nieto~\cite{braaten}, for a thorough
discussion of this approach and calculational details.  \\ \\
If we denote the scalar field in the full theory
by $\Phi$, the Euclidean Lagrangian for
massless $g^2\Phi^4$-theory is
\begin{equation}
{\cal L}_{E}=\frac{1}{2}(\partial_{\mu}\Phi )^{2}+\frac{g^2}{24}\Phi^{4}.
\end{equation}
In the effective theory, we similarly denote the scalar field by $\phi$.
The field in the three-dimensional Lagrangian can be approximately
identified with the light mode in the underlying theory. 
At leading order in $g^2$, the relation between them is
\beq
\label{rel}
\phi=\frac{1}{\sqrt{T}}\Phi.
\eeq
First, we must identify the symmetries of the  effective Lagrangian.
We have a Z$_2$-symmetry 
$\phi\rightarrow -\phi$, which follows from the corresponding symmetry in
the full theory. There is also a three-dimensional rotational symmetry.
Hence, we may write
\beq
{\cal L}_{\mbox{\footnotesize eff}}=
\frac{1}{2}(\partial_i\phi)^2+\frac{1}{2}m^2(\Lambda)\phi^2
+\frac{\lambda(\Lambda)}{24}\phi^4
+\frac{h_1(\Lambda)}{6!}\phi^6+h_2(\Lambda)
\phi^2\nabla^2\phi^2+\delta{\cal L}.
\eeq
Here, we have indicated that the parameters in the effective Lagrangian
generally depend on the ultraviolet cutoff $\Lambda$  in the
effective theory\footnote{The field $\phi$ also depends on the scale
$\Lambda$, but it is suppressed for notational ease.}. This is necessary
in order to cancel the $\Lambda$-dependence that arises in the calculations
using the three-dimensional effective
Lagrangian. According to Ref.~\cite{braaten},
$\Lambda$ can be viewed as an arbitrary factorization scale, which separates
the scales $T$ and $gT$.
Moreover, $\delta{\cal L}$ includes all other operators which
are consistent with the symmetries.
The parameters $m^2(\Lambda)$ and $\lambda (\Lambda)$ are the only operators
which contribute to the screening mass squared through order $g^5$.
Thus for the present calculation, these are the only quantities which
must be determined.

The parameters in the effective theory are determined by a matching 
requirement. We demand that static correlators in the full theory be the
same as those in the effective theory to some desired accuracy
at long distances $R\gg 1/T$. We shall carry out the
matching using strict perturbation 
theory~\cite{braaten}. 
Strict perturbation corresponds to the following partition of 
the Lagrangian into a free piece and an interacting part:
\bqa 
\label{sp1}
{\cal L}_{0}&=&\frac{1}{2}(\partial_{\mu}\Phi )^2,
\\ 
\label{sp2}
{\cal L}_{\mbox{\footnotesize int}}&=&
\frac{g^2}{24}\Phi^4.
\eqa
The Lagrangian of the effective theory is split likewise:
\bqa
\label{sp3} 
({\cal L}_{\mbox{\footnotesize eff}})_{0}&=&\frac{1}{2}(\partial_{i}\phi )^2,
\\ 
\label{sp4} 
({\cal L}\mbox{\footnotesize eff})_{\mbox{\footnotesize int}}
&=&\frac{1}{2}m^2(\Lambda)\phi^2+
\frac{\lambda (\Lambda)}{24}\phi^4.
\eqa
It is clear from Eqs.~(\ref{sp1}) and~(\ref{sp2}) that strict perturbation
theory is ordinary perturbation theory, and it is therefore afflicted 
with severe infrared divergences, since we have not rearranged our
Lagrangian in order to screen them.
Nevertheless, this does not prevent us from determining the 
parameters using this partition of the Lagrangians. 
The point is that the parameters in ${\cal L}_{\mbox{\footnotesize eff}}$
encode the physics on the 
scale $T$ and 
are insensitive to the scale $gT$. So as long as we make the same incorrect
assumptions in the effective theory, the divergences cancel in
the matching procedure and we can use strict perturbation theory to
compute the coefficients in ${\cal L}_{\mbox{\footnotesize eff}}$.
This implies that we are free to use any suitable
infrared cutoff. In the present work we use dimensional regularization.
Finally, 
since strict perturbation theory is ordinary perturbation theory,
it is clear that the parameters in the effective Lagrangian 
can be written as powers series in $g^2$.

When we match static Greens functions in the two theories, it is necessary
to take into account the different normalizations of the fields.
The simple relation Eq.~(\ref{rel}) breaks down beyond leading order in 
$g^2$, and this is due to the wave function renormalization of $\Phi$
in the full theory. 
In $g^2\Phi^4$-theory, this is a two-loop effect and so is proportional to
$g^4$. It is therefore relevant for calculations first at order $g^6$,
and so we can use Eq.~(\ref{rel}) as it stands.

\begin{figure}[htb]
\begin{center}
\mbox{\psfig{figure=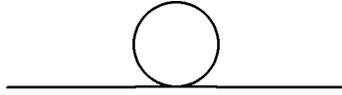}}
\end{center}
\caption{\protect One-loop correction to the two-point function.}
\label{1l}
\end{figure}

\begin{figure}[htb]
\begin{center}
\mbox{\psfig{figure=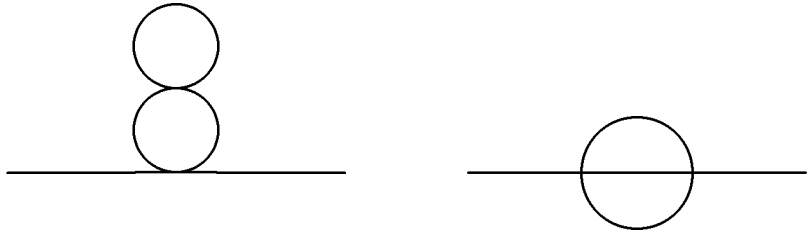}}
\end{center}
\caption{\protect Two-loop corrections to the two-point function.}
\label{2l}
\end{figure}

The mass parameter $m^2(\Lambda)$ must be determined at next-to-leading
order in $g^2$
Denoting the self-energy function of $\Phi$ by 
$\Sigma(k_0,{\bf k})$, we may write the static
two-point function in the full theory
as
\beq
\label{g2}
\Gamma^{(2)}(k_0=0,{\bf k})=k^2+\Sigma(k_0=0,{\bf k}).
\eeq 
Here, $k=|{\bf k}|$.
In Figs.~\ref{1l} and~\ref{2l} we have displayed the one and two-loop 
diagrams contributing to the self-energy function. The setting sun graph
is dependent on the external momentum $(k_0,{\bf k})$, but it is consistent
to evaluate this diagram at vanishing external momentum:
\beq
\label{sigma}
\Sigma (k_0=0,{\bf k}=0)=\frac{Z_{g^2}g^2}{2}
\hbox{$\sum$}\!\!\!\!\!\!\int_P\frac{1}{P^2}
-\frac{g^4}{4}\hbox{$\sum$}\!\!\!\!\!\!\int_{PQ}\frac{1}{P^2Q^4}
-\frac{g^4}{6}\hbox{$\sum$}\!\!\!\!\!\!\int_{PQ}\frac{1}{P^2Q^2(P+Q)^2}.
\eeq
Here, $Z_{g^2}$ is the renormalization constant for the coupling $g^2$.
The second term in Eq.~(\ref{sigma}) has a linear infrared divergence
for $q_0=0$, while the third term has a logarithmic infrared divergence
for $p_0=q_0=0$. Both singularities are of course canceled by 
corresponding IR-divergences in the effective theory.

\begin{figure}[htb]
\begin{center}
\mbox{\psfig{figure=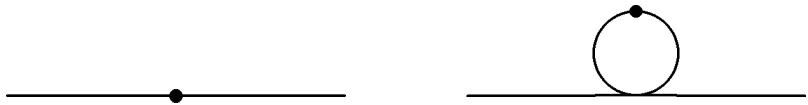}}
\end{center}
\caption{\protect Diagrams with a mass insertion contributing to
the two-point function in the effective theory.}
\label{dm}
\end{figure}

Let us now turn to the self-energy function of the field $\phi$, which
we denote by $\Sigma_{\mbox{\footnotesize eff}}({\bf k},\Lambda)$.
It has a perturbative expansion which is given by the diagrams 
in Figs.~\ref{1l} and~\ref{2l}, as well as the diagrams with a
mass insertion. These are depicted in Fig~\ref{dm}. 
The self-energy function of the field $\phi$ is also
evaluated at ${\bf k}=0$. This implies that
$\Sigma_{\mbox{\footnotesize eff}}(0,\Lambda)$ 
is identically zero, since there is no mass scale in the corresponding
integrals.
The two-point function in the effective theory can then be written as
\beq
\label{g2eff}
\Gamma^{(2)}_{\mbox{\footnotesize eff}}({\bf k},\Lambda)=k^2+m^2(\Lambda)+
\delta m^2.
\eeq
Here, we have included a mass renormalization counterterm $\delta m^2$.
The mass parameter $m^2(\Lambda)$ is then found by demanding 
\beq
\Gamma^{(2)}(k_0=0,{\bf k})=\Gamma^{(2)}_{\mbox{\footnotesize eff}}({\bf k},\Lambda).
\eeq
This implies that the mass parameter is given by 
\beq
\label{m2f}
m^2(\Lambda)=
\frac{Z_{g^2}g^2}{2}
\hbox{$\sum$}\!\!\!\!\!\!\int_P\frac{1}{P^2}
-\frac{g^4}{4}\hbox{$\sum$}\!\!\!\!\!\!\int_{PQ}\frac{1}{P^2Q^4}
-\frac{g^4}{6}\hbox{$\sum$}\!\!\!\!\!\!\int_{PQ}\frac{1}{P^2Q^2(P+Q)^2}-
\delta m^2.
\eeq
The necessary sum-integrals have been listed in Appendix A.
After renormalization of the coupling constant in Eq.~(\ref{m2f}), 
which is carried out
by the substitution
\beq
Z_{g^2}=1+\frac{3g^2}{32\pi^2\epsilon},
\eeq
we are still left with a pole in $\epsilon$.
This pole is canceled by $\delta m^2$ which reads
\beq
\delta m^2=\frac{g^4T^2}{24(4\pi)^2\epsilon}.
\eeq 
The mass parameter then becomes
\beq
\label{mass}
m^2(\Lambda)=\frac{g^2T^2}{24}\Bigg\{1+\frac{g^2}{(4\pi)^2}
\Big[-3\ln\frac{\mu}{4\pi T}+4\ln\frac{\Lambda}{4\pi T}-\gamma_E+2+
2\frac{\zeta^{\prime}(-1)}{\zeta (-1)}\Big]\Bigg\}.
\eeq
The mass parameter $m^2(\Lambda)$ was first obtained by 
Braaten and Nieto in Ref.~\cite{braaten}.
We have used the renormalization group equation for the
coupling $g^2$ 
\beq
\frac{dg^2}{d\mu}=\frac{3g^4}{16\pi^2},
\eeq
to change the scale from $\Lambda$ to $\mu$ and so the coupling constant
$g$ is then
evaluated at the scale $\mu$. 

It is clear from Eq.~(\ref{mass}) that $m^2(\Lambda)$
depends explicitly on the arbitrary scale
$\Lambda$. The mass parameter satisfies an evolution equation, which
can be derived from the requirement that the physical screening mass
be independent of this arbitrary scale $\Lambda$~\cite{braaten}.
Alternatively, one can differentiate Eq.~(\ref{mass}) 
with respect to $\Lambda$. In terms of $\lambda (\Lambda)$ the evolution
equation reads
\beq
\label{evol}
\Lambda\frac{dm^2}{d\Lambda}=\frac{1}{6}\Big(\frac{\lambda}{4\pi}\Big )^2.
\eeq

The physical interpretation of the mass parameter is that it is contribution
from the scale $T$ to the screening mass.

\begin{figure}[htb]
\begin{center}
\mbox{\psfig{figure=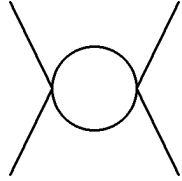}}
\end{center}
\caption{\protect One-loop correction to the four-point function.}
\label{q4}
\end{figure}

Let us now move on to the quartic self-interaction. We need
to know the coefficient $\lambda (\Lambda)$
in front at next-to-leading order in $g^2$.
Using similar arguments as we did when we computed the mass parameter,
one finds  
\beq
\lambda (\Lambda) =Z_{g^2}g^2T
-\frac{3}{2}g^4T\hbox{$\sum$}\!\!\!\!\!\!\int_P\frac{1}{P^4}.
\eeq
The first term on the right hand side results from matching at tree level,
while the second term arises from matching at the one-loop level. The
corresponding diagram is displayed in Fig.~\ref{q4}.
After renormalization of $g^2$, we find
\beq
\lambda(\Lambda) =g^2T-\frac{3g^4T}{(4\pi)^2}\Big[\ln
\frac{\Lambda}{4\pi T}+\gamma_{E}\Big].
\eeq
This coefficient was first found by Landsman in Ref.~\cite{lands} 
by explicitly
integrating out the nonzero Matsubara modes.
Using the renormalization group equation for $g^2$ one can demonstrate
that the quartic
coupling in the effective theory is independent of the scale $\Lambda$, and
so we can trade it for an arbitrary scale $\mu$.
\section{The Screening Mass}
In this section we calculate the screening mass squared to order $g^5$
using the effective Lagrangian.
When we calculated the parameters $m^2(\Lambda)$ and $\lambda(\Lambda)$, we
treated the infrared divergences incorrectly by decomposing our Lagrangians
according to Eqs.~(\ref{sp1}) and~(\ref{sp2}), and 
Eqs.~(\ref{sp3}) and~(\ref{sp4}).
In order to treat these divergences correctly, we must now incorporate
the infrared cutoff provided by the mass parameter
into the free part of ${\cal L}_{\mbox{\footnotesize eff}}$. 
Thus, we split
the effective Lagrangian according to
\bqa 
({\cal L}_{\mbox{\footnotesize eff}})_{0}&=&\frac{1}{2}(\partial_{i}\phi )^2+\frac{1}{2}m^2(\Lambda)\phi^2,
\\ 
({\cal L}_{\mbox{\footnotesize eff}})_{\mbox{\footnotesize int}}&=&
\frac{\lambda (\Lambda)}{24}\phi^4.
\eqa
The screening mass $m_s$
is defined as the pole of the propagator at spacelike
momentum~\cite{braaten}:
\beq
\label{propdef}
m_s^2=m^2(\Lambda)+\Sigma_{\mbox{\footnotesize eff}}({\bf k},\Lambda)
,\hspace{1cm}k^2=-m^2_s.
\eeq

\begin{figure}[htb]
\begin{center}
\mbox{\psfig{figure=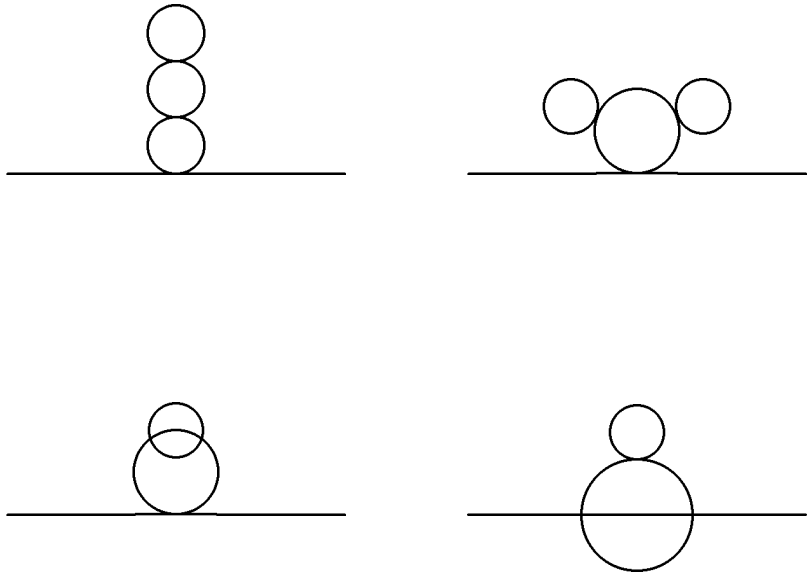}}
\end{center}
\caption{\protect Three-loop corrections to the two-point function
in the effective theory.}
\label{3l}
\end{figure}

The Feynman diagrams contributing to the self-energy function
$\Sigma_{\mbox{\footnotesize eff}}({\bf k},\Lambda)$ 
at next-to-next-to-leading order
are depicted in Figs.~\ref{1l} $-$~\ref{dm} and~\ref{3l}:
\bqa\nonumber
\label{3loop}
\Sigma_{\mbox{\footnotesize eff}}({\bf k},\Lambda)
&=&\frac{\lambda}{2}\int_p\frac{1}{p^2+m^2}-
\delta m^2\frac{\lambda}{2}
\int_p\frac{1}{(p^2+m^2)^2}
-\frac{\lambda^2}{4}\int_{pq}
\frac{1}{(p^2+m^2)^2(q^2+m^2)}-\\ \nonumber
&&\frac{\lambda^2}{6}\int_{pq}\frac{1}{(p^2+m^2)(q^2+m^2)
[({\bf p}+{\bf q}+{\bf k})^2+m^2]}+\\ \nonumber
&&\frac{\lambda^3}{8}
\int_{pqr}\frac{1}{(p^2+m^2)^2}\frac{1}{(q^2+m^2)^2}\frac{1}{(r^2+m^2)}+
\\ \nonumber
&&
\frac{\lambda^3}{8}\int_{pqr}\frac{1}{(p^2+m^2)^3(q^2+m^2)(r^2+m^2)}+\\ 
\nonumber
&&
\frac{\lambda^3}{8}\int_{pqr}\frac{1}{(p^2+m^2)^2(q^2+m^2)
[({\bf p}+{\bf q}+{\bf k})^2+m^2](r^2+m^2)}+\\ 
&&
\frac{\lambda^3}{12}\int_{pqr}\frac{1}{(p^2+m^2)(q^2+m^2)^2(r^2+m^2)
[({\bf p}+{\bf q}+{\bf r})^2+m^2]}+\delta m^2.
\eqa
The mass counterterm $\delta m^2$ cancels the pole in $\epsilon$ from the
second two-loop diagram in Fig.~\ref{2l}. Similarly, the divergence of the
one-loop diagram with a mass counterterm insertion cancels against a pole 
in $\epsilon$ arising from the third three-loop graph of Fig.~\ref{3l}.
Moreover, this one-loop diagram also gives finite contributions
to the screening mass, when $1/\epsilon$ ``hits'' the terms proportional
to $\epsilon$.
Finally,
the last two-loop diagram and the last three-loop diagram depend on the
external momentum ${\bf k}$. The effective expansion parameter of the 
quartic interaction is $\lambda/m$, and in order to 
compute the screening mass squared 
consistently in powers of $\lambda /m$, we must evaluate these integrals
at $k=im$. In order to see that this in fact is necessary, one
can perform an expansion in the external momentum $k$, and verify that all
terms are equally important for soft $k\sim gT$~\cite{parw1}.

The integrals appearing in the expression for the screening mass are tabulated
in Appendix B. Expanding the mass parameter in powers of $g^2$, we obtain
the screening mass squared through order $g^5$:
\bqa\nonumber
\label{ms1}
m^2_s&=&\frac{g^2}{24}T^2\Bigg\{1-\frac{g}{4\pi}\sqrt{6}+
\frac{g^2}{(4\pi)^2}
\Big[-3\ln\frac{\mu}{4\pi T}+4\ln\frac{2m}{4\pi T}-1+8\ln2-\gamma_E
+2+\\ \nonumber
&&2\frac{\zeta^{\prime}(-1)}{\zeta (-1)}\Big]
+\frac{g^3}{(4\pi)^3}\sqrt{6}\Big[\frac{9}{2}\ln\frac{\mu}{4\pi T}-
2\ln\frac{2m}{4\pi T}-3
-7\ln2+\frac{7}{2}\gamma_{E}- \\
&&\frac{\zeta^{\prime}(-1)}{\zeta (-1)}
 \Big]\Bigg\}.
\eqa
Firstly, one notes that the $\Lambda$-dependence explicitly
cancels in Eq.~(\ref{ms1}) to next-to-next-to-leading order in $g$.
Using the running of the coupling $g^2$, it also easy to verify that there
is no dependence on the renormalization scale $\mu$ either.
The result to order $g^3$ was first obtained by Dolan and Jackiw~\cite{dolan},
while Braaten and Nieto computed the screening mass squared to order $g^4$.
Moreover, they used the evolution equations, which the parameters in 
the effective
theory satisfy, to sum up leading logarithms of higher order in the 
perturbation expansion. The
result includes in particular a term proportional to $g^5\ln g$. This term
is of course also present in Eq.~(\ref{ms1}) above.
The complete result to order $g^5$ is new.

We would also like to point out that the effective field theory approach
explains the very appearance of these logarithms.
The mass parameter, and possibly other operators as well, depend explictly
on the factorization scale $\Lambda$ and such terms occur as logarithms
of $\Lambda/T$. In the effective theory, logarithms of $m/\Lambda$ arise
in perturbative calculations.
In order to cancel the $\Lambda$-dependence in physical
qauntities, these logarithms must match, leaving logarithms of $T/m$ or
$T/(gT)$. Hence, these logarithms can be attributed to the renormalization
of the parameters in the effective Lagrangian~\cite{braaten}, and in 
the present case it is the renormalization of the mass parameter.

The evolution eqution for the mass parameter, Eq.~(\ref{evol}),
can be used to sum up 
the leading logarithms of $T/(gT)$. This can be done by  
choosing the renormalization
scale $\mu=2\pi T$ and the factorization scale 
$\Lambda =\frac{gT}{2\sqrt{6}}$ in the expression for the 
mass parameter $m^2$~\cite{braaten}. 
The screening mass squared again follows from Eqs.~(\ref{propdef}) 
and~(\ref{3loop}), and in terms of $m^2$ and $\lambda$ it reads
\bqa
\label{ms2}
m^2_s&=&m^2\Bigg\{1-\frac{1}{2}(\frac{\lambda}{4\pi m})-
(\frac{\lambda}{4\pi m})^2[\frac{1}{8}-\frac{1}{2}\ln2]-
(\frac{\lambda}{4\pi m})^3[\frac{4}{96}+\frac{11}{96}\ln2]\Bigg\}.
\eqa
This is the full result to order $g^5$ and also includes leading logarithms
in the form $g^{2n+3}\ln^ng$, where $n$ is a natural number. 
These terms are obtained by expanding the
second term in Eq.~(\ref{ms2}) in powers of $g^2$. 
\section{Summary}
In the present work we have obtained the screening mass squared to order 
$g^5$ using effective field theory methods. 
The short-distance coefficients contains the physics on the scale $T$, while
the effective theory takes care of contributions to physical quantities
from the scale $gT$. Thus, effective field theory methods
unravel the contributions to physical quantities from the different
momentum scales, and streamlines calculations, since we treat one scale
in the problem at a time.
This is the advantage of the effective field approach
over the more conventional
resummation procedure; the latter complicates calculations  
unnecessarily because the sum-integrals involve both scales $T$ and $m$.

We were also able to sum up leading logarithms of $T/(gT)$ arising
from higher orders 
of perturbation theory by using the evolution equation satisfied by 
the mass parameter.

Our result is correct up to corrections of order $g^6$. In order to 
obtain the screening mass squared to this order, we must determine the
mass parameter to three-loop order and also compute the four-loop diagrams
in the effective theory. There are no new coefficients in the effective
Lagrangian which must be determined, since the corresponding operators
first contribute to $m^2_s$ at order $g^7$ or higher.
The calculation should be a
manageable task using the machinery developed to handle
difficult multi-loop sum-integrals~\cite{arnold1}, and complicated 
integrals in the effective theory.


\appendix\bigskip\renewcommand{\theequation}{\thesection.\arabic{equation}}
\setcounter{equation}{0}\section{Sum-integrals in the Full Theory}
In this appendix we give expressions for 
the sum-integrals used in the present work. 
We use the imaginary 
time formalism, where the four-momentum is $P=(p_{0},{\bf p})$
with $P^{2}=p_{0}^{2}+{\bf p}^{2}$. 
The Euclidean energy takes on discrete values, $p_{0}=2\pi nT$
for bosons.
Dimensional regularization is used to
regularize both infrared and ultraviolet divergences by working
in $d=4-2\epsilon$ dimensions, 
and we apply the $\overline{\mbox{MS}}$ 
renormalization scheme. 
We use the following shorthand notation
for the sum-integrals that appear below:
\bqa
\hbox{$\sum$}\!\!\!\!\!\!\int_P f(P)&\equiv 
&\Big( \frac{e^{\gamma_{\tiny E}}\mu^{2}}
{4\pi}\Big )^{\epsilon}\,\,\,\,
T\!\!\!\!\!
\sum_{p_{0}=2\pi nT}\int\frac{d^{3-2\epsilon}k}
{(2\pi)^{3-2\epsilon}}f(P).
\eqa
Then $\mu$ coincides with the
renormalization scale in the $\overline{\mbox{MS}}$ renormalization
scheme.

Arnold and Zhai have developed the machinery necessary to evaluate complicated
multi-loop sum-integrals~\cite{arnold1}. They have calculated and listed
the specific sum-integrals needed in the present work, and details
may be found in Ref.~\cite{arnold1}. We list them here for the convenience
of the reader 
\bqa
\hbox{$\sum$}\!\!\!\!\!\!\int_P\frac{1}{P^{2}}&=&\frac{T^{2}}{12}
\Big[1+\Big(2\ln\frac{\mu}{4\pi T}+
2+2\frac{\zeta^{\prime}(-1)}{\zeta (-1)}\Big)\epsilon
+O(\epsilon^{2})\Big],\\
\hbox{$\sum$}\!\!\!\!\!\!\int_P\frac{1}{(P^{2})^{2}}&=&\frac{1}{(4\pi )^{2}}
\Big[\frac{1}{\epsilon}+2\ln\frac{\mu}{4\pi T}
+2\gamma_{E}+O(\epsilon )\Big],\\
\hbox{$\sum$}\!\!\!\!\!\!\int_{PQ}\frac{1}{P^{2}Q^{2}(P+Q)^{2}}&=&0.
\eqa
\setcounter{equation}{0}\section{Integrals in the Effective Theory}
In the effective three-dimensional theory we use dimensional regularization
in $3-2\epsilon$ dimensions to regularize infrared and ultraviolet 
divergences.
In analogy with Appendix A, we define
\beq
\int_{p}f(p)\equiv\Big( \frac{e^{\gamma_{\tiny E}}\mu^{2}}
{4\pi}\Big )^{\epsilon}\int\frac{d^{3-2\epsilon}p}
{(2\pi)^{3-2\epsilon}}f(p).
\eeq
Again $\mu$ coincides with the renormalization scale in the 
modified minimal subtraction renormalization
scheme. 
All the integrals necessary in the present work, 
except for those in Eqs.~(\ref{e1}), ~(\ref{e2}) and
~(\ref{e3}) below, have been calculated by Braaten and Nieto in 
Ref.~\cite{braaten}. I have computed these remaining integrals 
using their methods. 
\bqa
\int_{p} \frac{1}{p^{2}+m^{2}}&=&-\frac{m}{4\pi}
\Big[1+\Big(2\ln\frac{\mu}{2m}+2\Big)\epsilon
+O(\epsilon^{2})\Big],\\
\int_{p} \frac{1}{(p^{2}+m^{2})^{2}}&=&\frac{1}{8\pi m}
\Big[1+\Big(2\ln\frac{\mu}{2m}\Big)\epsilon
+O(\epsilon^{2})\Big],\\
\label{e1}
\int_{p} \frac{1}{(p^{2}+m^{2})^{3}}&=&-\frac{1}{16\pi m^3}
\Big[1+\Big(2\ln\frac{\mu}{2m}+2\Big)\epsilon\Big], \\ \nonumber
\label{dimjens}
\hspace{-1cm}\lefteqn{\left.\int_{pq} \frac{1}{(p^{2}+m^{2})(q^{2}
+m^{2})[({\bf p}+{\bf q}+{\bf k})^{2}+m^{2}]}\right|_{k=im}}\hspace{1.5in} \\ 
&=&
\frac{1}{(8\pi )^{2}}\Big[\frac{1}{\epsilon}
+6
+4\ln \frac{\mu}{2m}-8\ln 2+O(\epsilon )\Big],\\ \nonumber
\label{e2}
\lefteqn{\left.\int_{pq} \frac{1}{(p^{2}+m^{2})^2(q^{2}
+m^{2})[({\bf p}+{\bf q}+{\bf k})^2+m^{2}]}\right|_{k=im}}
\hspace{1.5in}\\ 
&=&
\frac{1}{m^2(8\pi )^{2}}\Big[\ln2+O(\epsilon )\Big],\\ \nonumber
\label{e3}
\lefteqn{\int_{pqr} \frac{1}{(p^{2}+m^{2})^2(q^{2}
+m^{2})(r^2+m^2)[({\bf p}+{\bf q}+{\bf r})^{2}+m^{2}]}}\hspace{1.5in}\\ 
&=&\frac{1}{m(8\pi)^3}
\Big[\frac{1}{\epsilon}+6\ln\frac{\mu}{2m}-4\ln2+2+O(\epsilon )\Big].
\eqa

\newpage

\end{document}